\documentclass[10pt]{JHEP}
\usepackage{palatino}
\usepackage{amsmath}
\usepackage{amssymb}

\author{Joakim Munkhammar\\ Studentstaden 23:230, 752 33,
Uppsala\\
E-Mail: \email{joakim.munkhammar@gmail.com}}

\title{Linearization of Moffat's Symmetric Complex Metric Gravity}


\abstract{In this paper we investigate a complex symmetric
generalization of general relativity and in particular we
investigate its linearized field equations. We begin by reviewing
some basic definitions and structures in Moffat's symmetric
complex metric field theory of gravity. We then move on to derive
the linearized retarded complex field equations. In addition to
this we also derive a linearization of Moffat's field equations
based on the more rigorous Fermi coordinate approach. In
conclusion it is shown that the linearized symmetric complex field
equations leads to a complex form of gravitomagnetism. We also
briefly review the gravitational wave equation from the source
less linearized symmetric complex field equations and discuss some
open problems.}

\maketitle

\begin{document}

\section{Introduction}
General relativity is arguably the most successful theory of
physics in the 20:th century \cite{Unified}. A multitude of
attempts to modify general relativity in order to incorporate
other forces and quantum mechanics has therefore been carried out
\cite{Unified}. The complex metric generalization of general
relativity has an extensive history
\cite{Chamseddine2000p,Chamseddine2006p,Einstein1,Einstein2,Kao,Moffatp,Moffat56,Moffat57,Moffat572}.
In an attempt to unify gravity and electromagnetism Einstein
introduced the complex generalization of the metric tensor in 1945
\cite{Einstein1,Einstein2}. Recently this field of complex metric
gravity has drawn attention for a number of reasons
\cite{Chamseddine2000p,Chamseddine2006p,Moffatp}. One reason is
that there are promising leads to a quantum gravity in the complex
metric formalism and another is that there are signs within string
theory that the metric by necessity needs to be complex
\cite{Moffatp}. Although most approaches have regarded the
investigation of hermitian metric tensors there are some studies
that have focused on symmetric ones
\cite{Kao,Moffatp,Moffat56,Moffat57,Moffat572}. The first studies
on the possibility of a symmetric complex metric theory as a
unified field theory were done by Moffat
\cite{Moffat56,Moffat57,Moffat572}. The physical interpretation of
these symmetric complex metric field theories is still an open
problem and in this paper the symmetric complex field is
considered a modification of general relativity \cite{Moffatp}. In
this paper we shall study the linearized field of the symmetric
complex metric field theory proposed by Moffat
\cite{Moffat56,Moffat57}.




\section{Moffat's symmetric complex metric approach}
This symmetric complex metric field theory is based on Moffat's
approach \cite{Moffatp,Moffat56,Moffat57,Moffat572}. In general
the spacetime manifold is complex with coordinates (See section 3
\cite{Moffatp}):
\begin{equation}
z^\mu = x^\mu + i y^\mu,
\end{equation}
which can be generalized with hyperbolic complex coordinates in
order to avoid ghosts in the action \cite{Moffatp}. In that case
the field can be formulated as an 8-dimensional real spacetime
\cite{Moffatp}. In this approach the metric tensor has the form:
\begin{equation}
g'_{\mu \nu} = g_{\mu \nu} + i k \alpha_{\mu \nu},
\end{equation}
where $g_{\mu \nu}$ and $\alpha_{\mu \nu}$ are real-valued tensors
and $k$ is some real-valued constant and generally $'$ denotes
complex valued entities. The metric shall be considered symmetric:
\begin{equation}
 g_{\mu \nu} = g_{\nu \mu}, \hspace{20pt}
\alpha_{\mu \nu} = \alpha_{\nu \mu}, \hspace{20pt} g'_{\mu \nu} =
g'_{\nu \mu}.
\end{equation}
This gives the line element:
\begin{equation}\label{MasterMetric}
ds^2 = g'_{\mu \nu}dx^\mu dx^\nu = (g_{\mu \nu} + i k \alpha_{\mu
\nu}) dx^\mu dx^\nu.
\end{equation}
%
The component $g_{\mu \nu}$ is the \textit{gravitational metric
tensor} (equivalent to the one defined in general relativity) and
$\alpha_{\mu \nu}$ shall be referred to as the \textit{imaginary
metric tensor}. The definition of the covariant metric is given by
the usual definition (although the metric here is complex):
\begin{equation}\label{UpperLower}
g'_{\mu \nu} g'^{\mu \rho} = \delta_{\nu}^{\rho},
\end{equation}
as well as we shall require (In accordance with the approach by
Moffat \cite{Moffatp,Moffat56,Moffat57,Moffat572}):
\begin{equation}
g_{\mu \nu} g^{\mu \rho} = \delta_{\nu}^{\rho}.
\end{equation}
The definition of the field equations shall remain equivalent to
the Einstein field equations (EFE), apart from the fact that the
metric makes it complex:
\begin{equation}\label{Master}
R'_{\mu \nu} - \frac12 g'_{\mu \nu} R' = \frac{8 \pi G}{c^4}
T'_{\mu \nu}.
\end{equation}
These will be referred to as \textit{Moffat's symmetric complex
metric field equations} or simply \textit{the complex field
equations}. Here the Ricci tensor, Ricci scalar and the
stress-energy tensor are complex. The construction of the complex
metric field tensor \eqref{MasterMetric} can also be characterized
via complex vielbein $E'^{a}_{\mu} = Re[E'^{a}_{\mu}] + i
Im[E'^{a}_{\mu}]$:
\begin{equation}
g'_{\mu \nu} = E'^{a}_{\mu} E'^{b}_{\nu} \eta_{a b}.
\end{equation}
This type of construction has been used (with both real and
complex vielbeins) in order to reformulate general relativity as a
gauge theory \cite{Chamseddine2000p,Chamseddine2006p}. The
determination of Christoffel symbols follows from:
\begin{equation}
g'_{\mu \nu; \lambda} = \partial_{\lambda} g'_{\mu \nu} - g'_{\rho
\nu} \Gamma'^{\rho}_{\mu \lambda} - g'_{\mu \rho}
\Gamma'^{\rho}_{\nu \lambda} = 0.
\end{equation}
The complex Riemann tensor appears like:
\begin{equation}
R'^{\lambda}_{\mu \nu \sigma} = - \partial_\sigma
\Gamma'^{\lambda}_{\mu \nu} + \partial_\nu \Gamma'^{\lambda}_{\mu
\sigma} + \Gamma'^{\lambda}_{\rho \nu} \Gamma'^{\rho}_{\mu \sigma}
- \Gamma'^{\lambda}_{\rho \sigma} \Gamma'^{\rho}_{\mu \nu}
\end{equation}
and its contracted curvature tensor (Ricci tensor):
\begin{equation}
R'^{\sigma}_{\mu \nu \sigma} = R'_{\mu \nu}.
\end{equation}
Furthermore we may conclude that there will be four complex
(equivalent to eight real) Bianchi identities
\cite{Moffatp,Moffat56,Moffat57,Moffat572}:
\begin{equation}
(R'^{\mu \nu} - \frac12 g'^{\mu \nu} R')_{; \nu} = 0.
\end{equation}
Any vector transported in this complex metric field will have real
and imaginary components. In fact, if we look at the covariant
derivative of a complex vector $A'_\mu$ we get:
\begin{equation}\label{Transport1}
A'_{\mu; \nu} = A'_{\mu, \nu} - A'_\alpha \Gamma'^{\alpha}_{\mu
\nu},
\end{equation}
which, if we let $A'_\mu = V_\mu + i k A_\mu$ then
\eqref{Transport1} breaks up into the two equations:
\begin{align}
V_{\mu;\nu} = V_{\mu, \nu} + V_\alpha Re[\Gamma'^\alpha_{\mu \nu}]
- k A_\alpha Im[\Gamma'^{\alpha}_{\mu \nu}],\\
A_{\mu;\nu} = A_{\mu, \nu} + A_\alpha Re[\Gamma'^\alpha_{\mu \nu}]
+ V_\alpha Im[\Gamma'^{\alpha}_{\mu \nu}].
\end{align}
This shows that the transport of any real valued vector can result
in a complex valued one. Also it shows the interaction between
real and imaginary components of a vector along a transport.
Generally Moffat's complex field equations can be derived from a
complex Einstein-Hilbert action \cite{Moffat57,Moffat572}:
\begin{equation}\label{CompEinsteinHilbert}
S' = \frac{c^4}{16\pi G} \int d^4 x (\sqrt{-g'} g'^{\mu \nu}
R'_{\mu \nu}).
\end{equation}
It is a general belief that $S'$ has to be real under complex
metric circumstances
\cite{Einstein1,Kao,Moffatp,Moffat56,Moffat57,Moffat572}. Indeed,
if we apply the variational principle applied to the complex
Einstein-Hilbert action with matter term
\eqref{CompEinsteinHilbert} one gets the correct complex field
equations \cite{Moffat56,Moffat57,Moffat572}.

\section{Correspondence to real and imaginary gravity}
\subsection{Real gravity}
The real gravitational part becomes visible when the imaginary
tensor components $\alpha_{\mu \nu}$ vanishes:
\begin{equation}
g'_{\mu \nu} = g_{\mu \nu}.
\end{equation}
This scenario reduces the complex field theory \eqref{Master} to
being equivalent to the Einstein field equations of general
relativity:
\begin{equation}
R_{\mu \nu} - \frac12 g_{\mu \nu} R = \frac{8 \pi G}{c^4} T_{\mu
\nu},
\end{equation}
which are real valued.

\subsection{Imaginary gravity}
If we let the gravitational metric tensor $g_{\mu \nu} \to
\eta_{\mu \nu}$, which is the traditional vacuum situation in
general relativity, we get a form of imaginary gravity instead:
\begin{equation}
g'_{\mu \nu} = \eta_{\mu \nu} + i k \alpha_{\mu \nu}.
\end{equation}
This situation is not pure vacuum since the complex metric tensor
$\alpha_{\mu \nu}$ will still affect the background field. If we
in the most extreme case let $g_{\mu \nu} \to \eta_{\mu \nu}$ and
$\alpha_{\mu \nu} \to 0$ then we get a pure form of vacuum state
where no gravitational and no imaginary metric effects are
present:
\begin{equation}
g'_{\mu \nu} = \eta_{\mu \nu}.
\end{equation}
This field is governed by the Minkowski metric.

\section{Linearized symmetric complex field theory}

\subsection{Linearized retarded field equations}
This linearized retarded field equations approach was used by
Moffat in his original work on symmetric complex metric gravity
\cite{Moffat56,Moffat57,Moffat572}. Assume that the metric tensor
can be expressed as the Minkowski component and a linear
perturbation as:
\begin{equation}
g'_{\mu \nu} = \eta_{\mu \nu} + h'_{\mu \nu}.
\end{equation}
It is then useful to define the trace-reversed amplitude as:
\begin{equation}
\bar{h'}_{\mu \nu} = h'_{\mu \nu} - \frac12 \eta_{\mu\nu} h',
\end{equation}
where $h'=\eta^{\mu \nu}h'_{\mu \nu}$ is the trace of $h'_{\mu
\nu}$. The linearized version of the complex field equations
\eqref{Master} appears as follows:
\begin{equation}
\square \bar{h'}_{\mu \nu} = - \frac{16 \pi G}{c^4} T'_{\mu \nu},
\end{equation}
after imposing the Lorenz gauge condition $\bar{h'}^{\mu \nu}_{,
\nu} = 0$ \cite{Mashhoon1p}. If in addition to linearizing the
field equations assume that $|v|<<c$ it is useful to consider the
retarded solution as follows:
\begin{equation}
\bar{h'}_{\mu \nu} = \frac{4 G}{c^4} \int \frac{T'_{\mu
\nu}(ct-|\vec{x}-\vec{x}'|,\vec{x}')}{|\vec{x}-\vec{x}'|} d^3 x'.
\end{equation}
This lets us define $T'^{00} = \rho' c^2$ and $T^{0i} = c j'^i$,
with complex density and complex matter current defined as $\rho'$
and $\vec{j'}$ respectively. It is also useful here to define the
complex Gravitomagnetic (GM) potentials $(\Phi',\vec{A'})$ as
$\bar{h'}_{00} = 4 \Phi'/c^2$, $\bar{h'}_{0i} = -2A'_i/c^2$ and
$\bar{h'}_{ij} = O(c^{-4})$. The transverse gauge condition
implies:
\begin{equation}
\frac1c \frac{\partial \Phi'}{\partial t} + \frac12 \nabla \cdot
\vec{A'} = 0.
\end{equation}
Furthermore the wave equations of the $0i$-components become:
\begin{equation}
\square \Phi' = - 4 \pi G \rho',
\end{equation}
\begin{equation}
\square A'_i = \frac{8 \pi G}{c} j'^i.
\end{equation}
These equations make up a complex analogue of Maxwell's equations,
and are the complex gravitomagnetic field equations. This
linearized retarded approach is only valid in certain frames and
it does not bring the correct stress-energy tensor for neither
standard nor complex linearized gravity \cite{Mashhoon1}. This is
the primary reason why we shall address this more generally via a
Fermi coordinate approach below.

\subsection{Fermi coordinate approach to linear complex symmetric gravity}
In this section we are going to use Mashhoon's Fermi approach to
gravitomagnetism and apply it to the complex field equations
\eqref{Master}. In the local reference frame one may setup a Fermi
coordinate system along its path. This is equivalent to
constructing a an inertial system of coordinates in the immediate
neighborhood \cite{Mashhoon1}. One may then let
$\lambda^{\mu}_{(\alpha)}$ be the non-rotating orthonormal tetrad
of the reference observer (here taken to be real valued).
$\lambda^{\mu}_{(\alpha)}$ is a collection of unit vectors along
ideal gyro directions that are parallel transported along the
worldline \cite{Mashhoon1}. The new space will have Fermi
coordinates $X^\mu = (T,\vec{X})$. The Riemann tensor will be
projected on the orthonormal tetrad of the reference observer
according to:
\begin{equation}
R'_{\alpha \beta \gamma \delta} = R'_{\mu \nu \rho \sigma}
\lambda^{\mu}_{(\alpha)} \lambda^{\nu}_{(\beta)}
\lambda^{\rho}_{(\gamma)} \lambda^{\sigma}_{(\delta)}.
\end{equation}
This gives the space time metric:
\begin{align}
g'_{00} = -1 -R'_{0i0j}X^i X^j+..., \\
g'_{0i} = -\frac23 R'_{0jik} X^j X^k+...,\\
g'_{ij} = \delta_{ij} - \frac13 R'_{ikjl} X^k X^l+...
\end{align}
These coordinates are admissible within a cylindrical spacetime of
radius $\sim \textbf{R}$ (the radius of curvature of spacetime)
around the worldline of the reference observer \cite{Mashhoon1}.
This also means that $g'_{\mu \nu} = \eta_{\mu \nu}$ by
construction. Within this region it is possible to construct the
complex gravitomagntetic potentials:
\begin{align}\label{Fermi}
\Phi'(T,\vec{X}) =  -\frac12 R'_{0i0j}X^i X^j+..., \\
A'_i (T,\vec{X}) = \frac13 R'_{0jik} X^j X^k+...,
\end{align}
which in the real case is the expression for the four vector
$A_\mu$ of gravitomagnetism (GM). The setup \eqref{Fermi} allows
for the following construction of the complex GM fields:
\begin{align}
E'_i(T,\vec{X}) =  R'_{0i0j}X^j +..., \\
B'_i (T,\vec{X}) = -\frac12 \epsilon_{ijk} R'^{jk_{0l}}(T)
X^l+...,
\end{align}
which decomposes into the the general relativistic- and imaginary
parts (Here $E'_i$ and $B'_i$ are the complex
gravitoelectromagnetic fields, and $G_i$ combined with $B_{G i}$
are the real gravitoelectromagnetic fields):
\begin{align}
E'_i = G_i + i k E_i\\
B'_i = B_{G i} + i k B_i.
\end{align}
which allows for the construction of the complex Faraday tensor
(to linear order in $\vec{X}$):
\begin{equation}
F'_{\mu \nu} = \partial_\mu A'_\nu - \partial_\nu A'_\mu,
\end{equation}
which we can be decomposed into general relativistic and imaginary
tensor components:
\begin{equation}\label{Faraday1}
F'_{\mu \nu} = G_{\mu \nu} + i k F_{\mu \nu}.
\end{equation}
Also note that $F'_{0i} = -E'_i$ and $F'_{i j} = \epsilon_{i j k}
B'^k$. Then Maxwell's equations $F_[\alpha \beta, \gamma] = 0$ and
$F^{\alpha \beta}_{,\beta} = 4 \pi J^\alpha$ are satisfied to
lowest order in $|X|/\textbf{R}$ with:

\begin{equation}
4 \pi J'_\alpha(T,0) = - 8 \pi \Big(T'_{0\alpha}-\frac12 \eta_{0
\alpha} T'^\beta_\beta \Big)
\end{equation}

along the trajectory in Fermi coordinates \cite{Mashhoon1}. The
Lorenz force for the linearized complex field appears as
\cite{Mashhoon2}:
\begin{align}
\frac{\partial^2 X^i}{\partial T^2} + R'_{0i0j}X^j + 2R'_{ikj0}V^k
X^j + \big( 2R'_{0kj0}V^iV^k + \frac23 R'_{ikjl} V^k V^l + \frac23
R'_{0kjl} V^i V^k V^l \big) X^j = 0,
\end{align}
where $V^i = dX^i/dT$. For linear order velocity this becomes:
\begin{equation}
m \frac{\partial^2 \vec{X}}{\partial T^2} = - m \vec{E} - 2m
\vec{V} \times \vec{B},
\end{equation}
where $m$ is the mass of the test particle. This shows that the
entire linearized complex field is a spin-2 field, a remnant from
the spin-2 character of the complex field equations. The complex
stress-energy tensor in the Fermi approach can be set up along the
trajectory locally as:
\begin{align}\label{StressEnergy}
T'^{\alpha \beta} = \frac{1}{4 \pi G} \Big( F'^\alpha_\gamma
F^{\beta \gamma} - \frac14 g'^{\alpha \beta} F'_{\gamma \delta}
F'^{\gamma \delta} \Big) = \frac{1}{4 \pi G} \Big(
R'^\alpha_{\gamma 0 i} R'^{\beta \gamma_0 j} - \frac14
\eta^{\alpha \beta} R'_{\gamma \delta 0 i} R'^{\gamma \delta_{0
j}} \Big) X^i X^j.
\end{align}
This stress-energy tensor is only really valid when averaged over
a small domain though \cite{Mashhoon1,Mashhoon2}. The real parts
of the stress-energy tensor (for $g_{\mu \nu} \sim \eta_{\mu
\nu}$) becomes (together with the decomposition notation
\eqref{Faraday1}):
\begin{equation}
Re[T'^{\alpha \beta}] = \frac{1}{4 \pi G} \Big(
G^{\alpha}_{\gamma} G^{\beta \gamma} - \frac14 \eta^{\alpha \beta}
G_{\gamma \delta}G^{\gamma \delta} - k^2\Big(F^{\alpha}_{\gamma}
F^{\beta \gamma} - \frac14 \eta^{\alpha \beta} F_{\gamma
\delta}F^{\gamma \delta}\Big)\Big).
\end{equation}
Here the traditional gravitomagnetic part of the stress-energy
tensor is visible as the first two terms. Indeed, if we let $k \to
0$ or $\alpha_{\mu \nu} \to 0$ we get:
\begin{equation}
T'^{\alpha \beta} = T^{\alpha \beta} = \frac{1}{4 \pi G} \Big(
G^{\alpha}_{\gamma} G^{\beta \gamma} - \frac14 \eta^{\alpha \beta}
G_{\gamma \delta}G^{\gamma \delta} \Big).
\end{equation}
which is the traditional stress-energy tensor of gravitomagnetism
\cite{Mashhoon1,Mashhoon2}.

\subsection{Wave equations}
As an application of the linearized equations we shall here
construct the complex gravitational wave equation from the source
less wave equations (from \eqref{Fermi}):
%
\begin{align}\label{Wave}
\square A'_\mu = \square (A_{G \mu} + i k A_{E \mu}) = 0.
\end{align}
In the real valued situation equation \eqref{Wave} is a remnant
from the gravitomagnetic equations that appear in the linear
approach to general relativity \cite{Mashhoon2}. It has the
general solution:
\begin{equation}\label{AWave}
A'_\mu = c_\mu e^{i k_\nu x^\nu} = c_\mu (\cos(k_\nu x^\nu) + i
\sin(k_\nu x^\nu)),
\end{equation}
where $c_\mu$ and $k_{\nu}$ are a constant four vectors. In this
formalism the complex gravity radiation is always complex.





\section{Conclusions}
In this paper we have investigated Moffat's symmetric complex
metric generalization of general relativity. We reviewed some
basic results regarding the complex symmetric field as well as its
linearized retarded field equations. We showed that in the linear
case the complex symmetric field equations become a complex form
of gravitomagnetism (GM). As a more rigorous approach to GM we
used a Fermi coordinate approach. In connection with this we
derived complex gravity waves from the source less wave equations
arising in the complex gravitomagnetism equations. Many open
problems remain in complex symmetric gravity, like for example the
extension to spin fields and torsion via Einstein-Cartan theory.
Also the introduction of a non-symmetric complex field via Moyal
products could perhaps lead to a quantized version of the field
\cite{Moffatp}.


\acknowledgments I would like to thank prof. John Moffat at the
Perimeter Institute for valuable comments on this paper. I would
also like to thank Martin Th\"ornqvist for general discussions on
field theories.

\end{document}